\documentclass[prl,twocolumn,showpacs,superscriptaddress]{revtex4-1}

\usepackage{amsfonts,amsmath,amssymb,graphicx}
\usepackage{amsfonts}
\usepackage{amsmath}
\usepackage{amssymb}
\usepackage{graphicx}
\newcommand{\f}{\frac}
\begin{document}

\title{\textbf{Controlling the Momentum Current of an Off-resonant Ratchet} }
\author{R. K. Shrestha}
\affiliation{Department of Physics, Oklahoma State University, Stillwater, Oklahoma 74078-3072, USA}
\author{J. Ni}
\affiliation{Department of Physics, Oklahoma State University, Stillwater, Oklahoma 74078-3072, USA}
\author{W. K.  Lam}
\affiliation{Department of Physics, Oklahoma State University, Stillwater, Oklahoma 74078-3072, USA}
\author{S. Wimberger}
\affiliation{Institut f\"{u}r Theoretische Physik, Universit\"{a}t Heidelberg, Philosophenweg 19, 69120 Heidelberg, Germany}
\author{G. S. Summy}
\affiliation{Department of Physics, Oklahoma State University, Stillwater, Oklahoma 74078-3072, USA}

\begin{abstract}
\noindent We experimentally investigate the phenomenon of a quantum
ratchet created by exposing a Bose-Einstein Condensate to short pulses of a
potential which is periodic in both space and time. Such a ratchet
is manifested by a directed current of particles, even though there
is an absence of a net bias force. We confirm a recent theoretical
prediction [M. Sadgrove and S. Wimberger, New J. Phys. \textbf{11},
083027 (2009)] that the current direction can be controlled by
 experimental parameters which leave the underlying symmetries of the system unchanged. We demonstrate
that this behavior can be understood using a single variable containing many of the experimental parameters and thus the ratchet current is
describable using a single universal scaling law.
\end{abstract}

\maketitle
\section{I. INTRODUCTION}
Ever since the   realization of the atom optics quantum kicked rotor (AOQKR) \cite{aoqr}, it has been  one of the workhorses for studies of experimental quantum chaos. It has revealed a wide variety of interesting effects including: dynamical localization \cite{localize}, quantum resonances (QR) \cite{localize,ryu,fm}, quantum accelerator modes  \cite{fgr,gazal}, and quantum ratchets \cite{monterio,Ratchet,Ratchetp,racht,Ratcheta,Ratchets,Ratchetab,rachtheory,rocking}. The latter are quantum mechanical systems that display directed motion of particles  in the absence of unbalanced forces. They are of considerable interest because classical ratchets are the  underlying  mechanism for some  biological motors and nanoscale devices \cite{racht}.
Recent theoretical \cite{Ratcheta,Ratchetab} and experimental \cite{Ratchet} studies have demonstrated that a controllable directed current arises in  kicked atom systems at QR. A QR occurs when the kicking period is commensurate with the natural periods of the rotor and is characterized by a quadratic growth  of the kinetic energy with time. The question of what happens to a ratchet away from resonance was addressed in a recent theoretical paper \cite{njp}. In that work, the authors developed a classical-like ratchet theory and  proposed the existence of a one-parameter scaling law that could be used to predict  the ratchet current for a wide variety of parameters. It was also shown that an {\em {inversion}} of the momentum current is possible for some sets of scaling variables.

\begin{figure}[!h]
\includegraphics[width=9.5 cm, height=10 cm]{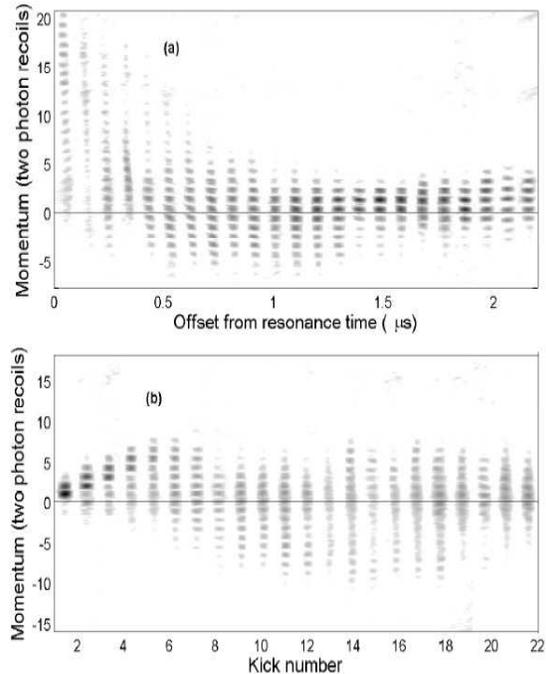}
\caption{Experimental momentum distributions after exposing a BEC to short pulses of an off-resonant standing wave of light. The momentum distributions are shown as a function of (a) pulse period offset from resonance ($\mu$s) (10 kicks, $\phi_d= 2.6$, $\gamma= -\pi/2$), and (b) kick number ($|\varepsilon| = 0.18$,  $\phi_d= 1.8$ and $\gamma= -\pi/2$.). Each momentum distribution was captured in a separate time-of-flight experiment. Note that there are features common to both panels, such as a weighting of the distributions towards positive momenta at small values of the independent variable followed by a tendency towards negative momenta at larger values of this parameter. These features are a manifestation of the fact that the mean momentum or ratchet current can be described by a universal scaling law. }\label{rawplot}
\end{figure}

In this paper, we report the experimental observation of such a ratchet current inversion and the verification of the scaling law for a wide variety of experimental parameters. Our experiments were carried out by  exposing a Bose-Einsten condensate (BEC) to a series of standing wave laser pulses that provided an optical potential periodic in space and time. Figure \ref{rawplot} shows raw  momentum distributions as a function of the pulse period's offset from the first QR  and the kick number (Fig. \ref{rawplot} (a) and (b) respectively).   It can be seen that there are certain values of time offset and kick number where the distribution is weighted more strongly towards negative momentum. This is evidence of a current reversal. Furthermore,  Fig. \ref{rawplot} (a) and (b) contain other similarities. For example, there are parameter regimes where the 
momentum distributions tend strongly towards positive momenta, followed by the current reversal regions where the distributions tend  negative. This suggests that it may be possible to use a single-parameter theory to understand the behavior of the system. Moreover, since the time offset from QR   effectively defines a new Planck constant \cite{fgr,WimNL}, we can easily switch from the classical  to the quantum regime by a simple change of the pulse period \cite{comment}.
\section{II. THEORY}
The dynamics of the AOQKR system can be described by a Hamiltonian which in dimensionless units is \cite{fgr,ishan,sadg}:  $\hat{H}=\frac{\hat{p}^{2}}{2}
+\phi_{d}\cos(\hat{X})\sum_{t=1}^{N}\delta (t'-t\tau)$, where $\hat{p}$ is the momentum (in units of  $\hbar G$, two photon recoils)
 that an atom of mass $M$ acquires from short, periodic pulses of a standing wave with a grating vector $G=2\pi/\lambda_{G}$
($\lambda_{G}$ is the spatial period of the standing wave). Since
momentum in this system is only changed in quanta of $\hbar G$, we
break down $p$ as $p=n+\beta$ where $n$ and $\beta$ are integer and
fractional parts of the momentum respectively and  $\beta$, the
quasi-momentum, is conserved. Other variables are the position
$\hat{X}$ (in units of $G^{-1}$), the continuous time variable $t'$
(integer units), and the kick number $t$. The pulse period $T$ is
scaled by $T_{1/2}=2\pi M/\hbar G^{2}$ (the half-Talbot time) to
give the scaled pulse period $\tau=2\pi T/T_{1/2}$. The strength of
the kicks is given by $\phi_{d}=\Omega^{2}\Delta t/8\delta_{L}$,
where $\Delta t$ is the pulse length, $\Omega$ is the Rabi
frequency, and $\delta_{L}$ is the detuning of the kicking laser
 from the atomic transition. To create a ratchet from this Hamiltonian
 it was shown in \cite{Ratchet} that a superposition of two
plane waves should be used for the initial state.

A successful approach to treating this system close to resonant values of $\tau$ (i.e. $\tau=2\pi l$, with $l>0$ integer) is the so called $\varepsilon-$classical theory. Here the scaled pulse period is written as $\tau= 2\pi l +\varepsilon$, where $|\varepsilon|\ll1$, and can be shown to play the role of Planck's constant. In this case the dynamics can be understood by the classical mapping \cite{fgr,sadg},
\begin{equation}\label{map}
  J_{t+1}=J_t+\tilde{k}\sin(\theta_{t+1}),\hspace{4mm} \theta_{t+1}=\theta_t+J_t,
\end{equation}
where $\tilde{k}=|\varepsilon|\phi_d$ is the scaled kicking strength, $J_t=\varepsilon p_t +l\pi +\tau \beta$ is the scaled momentum variable and $\theta_t= X \mod(2\pi) + \pi[1-\text{sign}(\varepsilon)]/2 $ is the scaled position exploiting the spatial periodicity of the kick potential. As mentioned above, for the ratchet we start with a superposition of plane waves $|\psi_0\rangle=\f{1}{\sqrt{2}}\left[|0 \hbar G\rangle +e^{i\gamma}|1 \hbar G\rangle\right]$, or equivalently a rotor state $\f{1}{\sqrt{4\pi}}[1+e^{i(\theta+\gamma)}]$. This leads to the position space probability distribution function  $P(\theta)=|\psi(\theta)|^2=\f{1}{2\pi}[1+\cos(\theta+\gamma)]$.  Here $\gamma$  is an additional phase used to  account for the possibility that the initial spatial atomic distribution is shifted in position relative to the applied periodic potential. Although the distribution $P(\theta)$ is quantum in origin, in what follows it will be interpreted as a classical probability.
\begin{figure}[htp]
\includegraphics[width=9.00cm, height=9.5cm ]{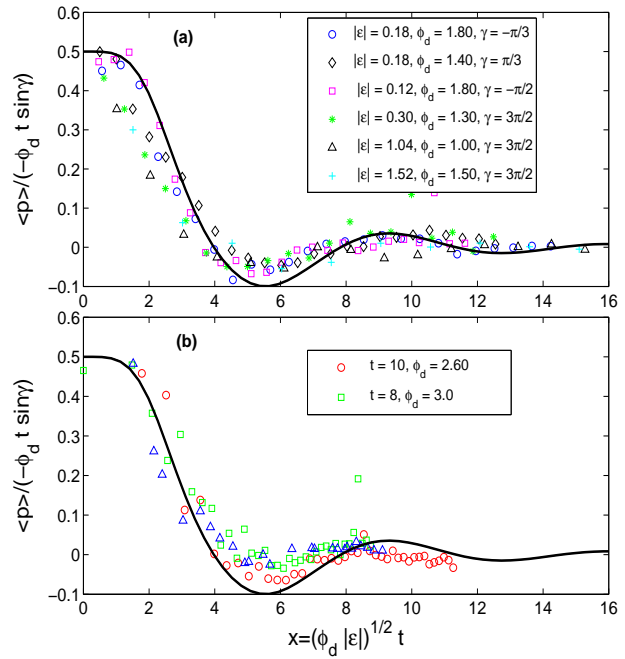}\caption{(Color online) Scaled mean momentum $\langle p \rangle/(-\phi_d t \sin\gamma)$ as a function of the scaling variable $x=\sqrt{(\phi_d |\varepsilon|)}$ $t$. In (a) $x$ was varied by scanning over kick number for different combinations of $\phi_d$, $|\varepsilon|$ and $\gamma$. In panel (b) $x$ was varied by scanning over $|\varepsilon|$ with $t=8$, $\phi_d = 3.0$ (green squares), and  with $t=10$, $\phi_d = 2.6$ (red circles). Also plotted in (b) is data from a scan  over $\phi_d$ with $|\varepsilon| = 0.18 $, $t = 8$ (blue triangles). In both panels, the solid line  is the function $F(x)/x$ given by Eq. (\ref{analytic}). This demonstrates that no matter how $x$ is obtained the scaled mean momentum is approximately  universal.}\label{scaleplot}
\end{figure}

 The original application of $\varepsilon-$classical theory to the kicked rotor system showed the existence of a one-parameter scaling law for the mean energy \cite{WimNL}. This was experimentally verified  in the vicinity of the first and second  quantum resonances ($l=1$ and $l = 2$) in Ref. \cite{mescaling}. It was found that the scaled energy could be written as $\f{E}{\phi_d^2 t}= 1-\Phi_0(x) + \f{4}{\pi x} G(x)$  where $x=\sqrt{\phi_d |\varepsilon|}$ $t$ is a scaling variable and $\Phi_0(x)$ and $G(x)$ are closed form functions of $x$. Recently, the existence of a one-parameter scaling law for the ratchet current using the same scaling parameter $x$ was proposed \cite{njp}. One of the notable features of this theory is that it predicts that at some values of the scaling variable (i.e. certain  families of real parameters) an inversion of momentum current can occur.

 In the pendulum approximation \cite{pendulum}, the motion of the kicked rotor in continuous time is described by the scaled Hamiltonian $H'\approx (J')^2/2 + |\varepsilon|\phi_d \cos(\theta)$. Here $J'=J/(\sqrt{\phi_d |\varepsilon|})$ is a scaled momentum variable. Near the quantum resonance, using the  position space probability distribution function $P(\theta)$, one can calculate $\langle J'-J_0'\rangle=\int_{- \pi}^{\pi}d\theta_0 P(\theta_0)(J'-J_0')$.  For $|\varepsilon| \lesssim 1$,  Eq. (\ref{map}) gives a phase space dominated by a pendulum-like resonance island of extension $4 \sqrt{\tilde k} \gg |\varepsilon|$ \cite{WimNL}. Hence $p=0$ and $p=1$ essentially contribute in the same way giving $J_0'=0$ so that  the map  in Eq. (\ref{map}) is
 $J'_{t+1}= \sqrt{\tilde{k}} \sum_{t=0}^{t=N-1}\sin(\theta_{t+1})$.  With the scaling variable $x$, the average scaled momentum becomes $ \langle J'-J_0'\rangle= -  \sin\gamma F(x)$, where

\begin{equation}\label{anas}
  F(x)= \f{1}{2\pi}\int_{-\pi}^{\pi} \sin\theta_0 J'(\theta_0, J_0'=0,x)d\theta_0.
\end{equation}
Thus the mean momentum (units of $\hbar G$) expressed  in terms of the scaled variables is
\begin{align}\nonumber\label{analytic}
  \langle p\rangle=\sqrt\f{\phi_d}{|\varepsilon|}\langle J'-J_0'\rangle &= -\f{\phi_d t \sin\gamma}{x}F(x)\\  \f{ \langle p\rangle}{{-\phi_d t \sin\gamma}}&= \f{F(x)}{x}
\end{align}
 where $F(x)$ can be computed from the above pendulum approximation \cite{njp}.
\section{III. EXPERIMENTS AND RESULTS}
 We performed our experiments  using a similar set up to that described in \cite{ishan}. A  BEC of  about 40000 $^{87}$Rb atoms was created
in the $5S_{1/2}$, $ F=1$ level using an all-optical trap technique. Approximately 5 ms   after  being   released from the trap, the condensate was exposed to a  pulsed  horizontal standing wave of wavelength $\lambda_G$. This was formed by two laser beams of wavelength
 $\lambda=$ 780 nm,  detuned $6.8 $GHz to the red of the atomic transition.
  The direction of each beam was aligned at $53^{\text{\textrm{o }}}$ to the vertical to give $\lambda_G=\lambda/(2\sin53^o)$. With these parameters the primary QR (half-Talbot time \cite {ryu,lepers,talbot}) occurred at multiples of $51.5 \pm 0.05$ $\mu $s. Each laser beam was  passed through an acousto-optic modulator  driven by an arbitrary waveform generator. This enabled  control of the phase,  intensity, and pulse length as well as the relative frequency between the kicking beams. Adding two counterpropagating waves differing in frequency by $\Delta f$ results in a standing wave that moves with a velocity $v=2\pi\Delta f/G$. The initial momentum or quasi-momentum $\beta$ of the BEC relative to the standing wave is proportional to $v$, so that by changing  $\Delta f$ the value of $\beta$ could be systematically controlled. The kicking pulse length was fixed at  1.54 $\mu$s, so   we varied the intensity rather than the pulse length to change the kicking strength  $\phi_d$.  This was done by adjusting the amplitudes of the RF waveforms driving the kicking
pulses, ensuring that the experiments were performed  in the Raman-Nath regime (the distance an atom
travels during the pulse is much smaller than the period of the potential).

The initial state for the experiment was prepared as a superposition of two momentum states $|p = 0 \hbar G\rangle $ and $|p = 1 \hbar G\rangle $ by applying a long ($\Delta t=38.6 \mu $s) and very weak standing wave pulse (Bragg pulse).   By using a  pulse of suitable  strength, an equal superposition of the two aforementioned atomic states was created ($\pi/2$ pulse).  The Bragg pulse was immediately followed by the kicking pulses in which  a relative phase  of $\gamma$ between the beams was applied. This phase was experimentally controlled by adjusting the phase difference between the RF waveforms driving the two AOMs. Finally the  kicked atoms were absorption imaged after $9$ ms using a time-of-flight measurement technique to yield momentum distributions like those seen in Fig. \ref{rawplot}.

We now discuss the experiments that were carried out to observe the ratchet effect away from \textbf{$l=1$} resonance. In this case $\beta=0.5$  is needed to fulfill the resonance condition \cite{Ratchet}. The measurements involve the determination of the  mean momentum of kicked BECs  for various combinations of the parameters $t$, $\phi_d$, $\varepsilon$ and $\gamma$. The measured momentum was then scaled by $- \phi_d t \sin\gamma$ and is plotted as a function of the scaling variable $x$ in Fig. \ref{scaleplot}. In Fig. \ref{scaleplot}(a) $x$ was changed by varying kick number, $t$, while in Fig. \ref{scaleplot}(b) different $x$ were obtained by scanning either $|\varepsilon|$ (red circles and green squares) or $\phi_d$ (blue triangles).   The solid line in both panels is a plot of the function $\f{F(x)}{x}$ given by Eq. (\ref{analytic}). It can be seen that no matter how $x$ is varied, the experimental results are in good agreement with the theory for many different combinations of parameters. In addition, there is a regime over  $x$  where an inversion of the ratchet current  takes place, with a maximum inversion  at $x \approx 5.6$. Interestingly this reversal of the ratchet takes place without altering any of the centers of symmetry of the system. Finally, it should be noted that the theory of Ref. \cite{njp} also predicts current inversions at higher values of $x$, although these were not seen in the experiments presumably because of dephasing effects such as vibrations and spontaneous emission. Even though the $\varepsilon-$classical theory assumes $|\varepsilon|$ is small, the experimental results show that it remains valid for higher values of $|\varepsilon|$ as well. In fact the window of valid $|\varepsilon|$  depends  on the kick number \cite{WimNL}, being rather large for small $t \lesssim 10-15$. This is expected from a Heisenberg/Fourier argument \cite{sadg,mescaling,masa}.
\begin{figure}[htp]
\includegraphics[width=8 cm, height=7 cm]{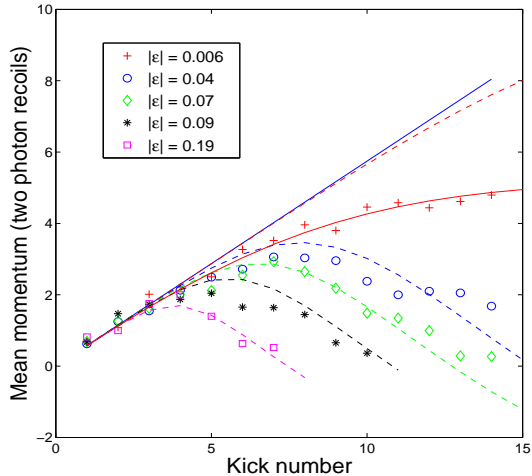}
\caption{(Color online) Momentum current as a function of kick number for  $|\varepsilon| = 0.006$ (red crosses), $|\varepsilon| = 0.04$ (blue circles), $|\varepsilon| = 0.07$ (green diamonds), $|\varepsilon| = 0.09$ (black stars) and  $|\varepsilon| =  0.19$ (purple squares). The blue solid line is the plot of  $\langle p_{t,res}\rangle = -\f{\phi_d t}{2}\sin\gamma $ for $\phi_d = 1.3$ and $\gamma = -\pi/3$. The dashed lines are the plot of Eq. (\ref{analytic}) with corresponding $|\varepsilon|$ and the red solid line is the plot of Eq.(2) in Ref. \cite{Ratchet} for $\beta = 0.5$ and $\Delta \beta = 0.02$.
}\label{momepsilon}
\end{figure}

We also investigated the sensitivity of the finite spread in initial quasi-momentum, $\Delta\beta$ to the momentum transfer.
 Figure \ref{momepsilon} shows the plot of momentum current as a function of kick number for different values of $|\varepsilon|$. The solid blue line is the plot of  $\langle p_{t,res}\rangle = -\f{\phi_d t}{2}\sin\gamma $.   The dashed lines are plots of Eq. (\ref{analytic}) for the corresponding experimental parameters.  The experimental results show that the
    farther one goes from  resonance the sooner the momentum current turns towards  negative values (current reversal).
    These results are in    good agreement with the theory except  very close to resonance,
    where the red dashed curve fits poorly to the red crosses . For this data,
    we note that  the suppression in momentum current is likely to be caused mainly by the effect of the initial spread of quasi-momentum.
    This phenomenon was also seen in Ref. \cite{Ratchet} where the ratchet current for
    finite $\Delta \beta$ was shown to be
     $ \langle p_{t,res}\rangle_{\Delta\beta}=\f{\phi_d}{2} \sum_{s=1}^{t}\sin[(\pi l+\tau\beta)s-\gamma]\exp[-2(\pi l \Delta\beta s)^2]$.
     This equation with $\Delta \beta =0.02$ (independently estimated from time-of-flight measurements) is also plotted in Fig. \ref{momepsilon}
      (red solid line). It can be seen to agree well with experiment. Thus  for $|\varepsilon|\gtrsim 0.04$  (corresponding to an offset from resonance of 0.3 $\mu$s), $\Delta\beta$ plays an
      unimportant role in the dynamics of the ratchet. This is because at resonance the total phase  the momentum states acquire must be an integer multiple of $2\pi$. Any deviation from this condition significantly suppresses the momentum current at longer times. However the momentum state phases away from resonance are already pseudo-random, so the phase changes caused by $\Delta\beta$ have a negligible effect.
\section{IV. CONCLUSIONS}
 We have performed experiments to  observe an
off-resonant atomic ratchet by exposing an initial atomic state  which was
a superposition of two momentum states to a series of standing wave pulses. We measured the mean momentum current as a function of a scaling variable $x$, which contained important pulse parameters such as the offset of the kicking period from resonance, the kick number, and the kick strength. We showed that a scaled version of the mean momentum could be described solely by $x$, a result postulated by a theory based on a classical treatment of the system \cite{njp}. The experiment verified that for certain ranges of $x$ the momentum current exhibited an inversion. We also studied
   the effect of initial quasi-momentum width on  the ratchet current away from  resonance. This  width has a  large impact extremely  close to resonance, but plays an unimportant role as we go only a  little farther from resonance. Ultimately one can now control the strength and direction of the ratchet without changing the underlying relative symmetry between the initial state and the potential.  This has practical advantages, since it is very easy to control the kicking strength, period, or kick number and hence influence the scaling parameter $x$.  Another interesting possibility is the investigation of the performance of $\varepsilon-$classical ratchet theory as $|\varepsilon|$ becomes larger. This should allow the crossover from classical to quantum ratchet dynamics \cite{crossover} to be better understood.
\section{ACKNOWLEDGEMENTS}
   This work was partially supported by the NSF under Grant No. PHY-0653494.
SW is very grateful to Mark Sadgrove for fruitful collaborations, for the cordial hospitality at OSU, and for support from the DFG through FOR760 (WI 3426/3-1), the HGSFP (GSC 129/1), the CQD and the Enable Fund of Heidelberg University. We also thank to I. Talukdar for helpful discussions.


\begin{thebibliography}{99}
\bibitem{aoqr} F. L. Moore,  J. C. Robinson, C. F. Bharucha, B. Sundaram, and M. G. Raizen, Phys. Rev. Lett. \textbf{75}, 4598 (1995).
\bibitem{localize} F. L. Moore, J. C. Robinson, C. Bharucha, P. E. Williams, and M. G. Raizen, Phys. Rev. Lett. \textbf{73}, 2974 (1994).
\bibitem{ryu} C. Ryu, M. F. Anderson, A. Vaziri, M. B. d$'$Arcy, J. M. Grossman, K. Helmerson, and W. D. Phillips, Phys. Rev. Lett. \textbf{96}, 160403 (2006).
\bibitem{fm} F. M. Izrailev, Phys. Rep. \textbf{196}, 299 (1990).
\bibitem{fgr} S. Fishman, I. Guarneri and L. Rebuzzini, Phys. Rev. Lett. \textbf{89}, 084101 (2002); J. Stat. Phys. \textbf{110}, 911 (2003).
\bibitem{gazal} G. Behinaein, V. Ramareddy, P. Ahmadi, and G. S. Summy, Phys. Rev. Lett. \textbf{97}, 244101 (2006); V. Ramareddy, G. Behinaein, I. Talukdar, P. Ahmadi and G. S. Summy, Eur. Lett. \textbf{89}, 33001 (2010); M. K. Oberthaler, R. M. Godun, M. B. d'Arcy, G. S. Summy, and K. Burnett, Phys. Rev. Lett. \textbf{83}, 4447 (1999).
\bibitem{monterio} T. S. Monteiro,  P. A. Dando, N.  Hutchings , M. Isherwood, Phys. Rev. Lett. \textbf{89}, 194102 (2002).
\bibitem{Ratchet}  I. Dana, V. Ramareddy, I. Talkukdar, and G. S. Summy,, Phys. Rev. Lett. \textbf{100}, 024103 (2008).
\bibitem{Ratchetp} M. Sadgrove, M. Horikoshi, T. Sekimura and K. Nakagawa, Eur. Phys. J. D  \textbf{45}, 229 (2007).
\bibitem{racht} P. Reimann, Phys. Rep.  \textbf{361}, 57 (2002); R. D. Astumian and P. H\"{a}nggi, Phys. Today  \textbf{55}, No.11, 33 (2002).
\bibitem{Ratcheta} M. Sadgrove,  M. Horikoshi, T. Sekimura, and K. Nakagawa, Phys. Rev. Lett. \textbf{99}, 043002 (2007).
\bibitem{Ratchets} T. Salger, S. Kling, T. Hecking, C. Geckeler, L. M.-Molina, M. Weitz, Science, \textbf{326}, 1241 (2009).

\bibitem{Ratchetab} I. Dana and V. Roitberg, Phys. Rev. E. \textbf{76}, 015201(R) (2007).
\bibitem{rachtheory} E. Lundh and M. Wallin, Phys. Rev. Lett. \textbf{94}, 110603 (2005).
\bibitem{rocking} A. Wickenbrock, D. Cubero, N. A. Abdul Wahab, P. Phoonthong, F. Renzoni, Phys. Rev. E \textbf{84}, 021127 (2011).
\bibitem{njp} M. Sadgrove and S. Wimberger,   New J. Phys.  \textbf{11}, 083027 (2009).
\bibitem{WimNL} S. Wimberger,  I. Guarneri and S. Fishman, Nonlinearity \textbf{16}, 1381 (2003).
\bibitem{comment} At QR there is effectively no evolution between the kicks, making the system formally equivalent to the situation where the pulse period and therefore the scaled Planck's constant are zero.
\bibitem{ishan} I. Talukdar, R. Shrestha and G. S. Summy, Phys. Rev. Lett. \textbf{105}, 054103 (2010).
\bibitem{sadg} M. Sadgrove  S. Wimberger, S. Parkins, and R. Leonhardt,    Phys. Rev. Lett. \textbf{94}, 174103 (2005).
\bibitem{mescaling}S. Wimberger,  M. Sadgrove, S. Parkins, and R. Leonhardt, Phys. Rev. A \textbf{71}, 053404 (2005).
\bibitem{pendulum} G. Casati and I. Guarneri, Commun. Math. Phys. \textbf{95}, 121 (1984).
\bibitem{lepers} M. Lepers, V. Zehnl\'e and J. C. Garreau, Phys. Rev. A \textbf{77},  043628 (2008).
\bibitem{talbot} L. Deng, E. W. Hagley, J. Denschlag, J. E. Simsarian, Mark Edwards, Charles W. Clark, K. Helmerson, S. L. Rolston, and W. D. Phillips,, Phys. Rev. Lett. \textbf{83},  5407 (1999).
\bibitem{masa} M. Sadgrove and S. Wimberger, Adv. At. Mol. Opt. Phys. \textbf{60}, 315 (2011).
\bibitem{crossover} H. Schanz, M.-F. Otto, R. Ketzmerick, and T. Dittrich, Phys. Rev. Lett. \textbf{87},  070601 (2001).
\end{thebibliography}
\end{document}